\documentclass[twocolumn,twocolappendix]{aastex7}
\usepackage{graphicx}
\usepackage{amsmath}
\usepackage{scalerel}
\usepackage{apjfonts}
\usepackage{tikz}
\usetikzlibrary{svg.path}
\usepackage{comment}
\usepackage{array,multirow,color,tablefootnote}
\usepackage{xcolor}

\definecolor{orcidlogocol}{HTML}{A6CE39}
\tikzset{
  orcidlogo/.pic={
    \fill[orcidlogocol] svg{M256,128c0,70.7-57.3,128-128,128C57.3,256,0,198.7,0,128C0,57.3,57.3,0,128,0C198.7,0,256,57.3,256,128z};
    \fill[white] svg{M86.3,186.2H70.9V79.1h15.4v48.4V186.2z}
                 svg{M108.9,79.1h41.6c39.6,0,57,28.3,57,53.6c0,27.5-21.5,53.6-56.8,53.6h-41.8V79.1z M124.3,172.4h24.5c34.9,0,42.9-26.5,42.9-39.7c0-21.5-13.7-39.7-43.7-39.7h-23.7V172.4z}
                 svg{M88.7,56.8c0,5.5-4.5,10.1-10.1,10.1c-5.6,0-10.1-4.6-10.1-10.1c0-5.6,4.5-10.1,10.1-10.1C84.2,46.7,88.7,51.3,88.7,56.8z};
  }
}

\newcommand\orcidicon[1]{\href{https://orcid.org/#1}{\mbox{\scalerel*{
\begin{tikzpicture}[yscale=-1,transform shape]
\pic{orcidlogo};
\end{tikzpicture}
}{|}}}}

\newcommand{\ms}{m\,s$^{-1}$}
\newcommand{\um}{{\textmu}m}
\newcommand{\persec}{s$^{-1}$}
\newcommand{\water}{H$_{2}$O}
\newcommand{\cotwo}{CO$_{2}$}
\newcommand{\methane}{CH$_{4}$}
\newcommand{\methanol}{CH$_{3}$OH}
\newcommand{\ethane}{C$_{2}$H$_{6}$}

\newcommand{\qwater}{Q_{\rm H_{2}O}}
\newcommand{\qcotwo}{Q_{\rm CO_2}}
\newcommand{\qmethane}{Q_{\rm CH_4}}
\newcommand{\qnickel}{Q_{\rm Ni}}
\newcommand{\trot}{T_{\rm rot}}

\begin{document}

\title{The Volatile Inventory of 3I/ATLAS as seen with JWST/MIRI}

\correspondingauthor{Matthew~Belyakov}
\author[orcid=0000-0003-4778-6170]{Matthew Belyakov}
\altaffiliation{Authors contributed equally to this work.}
\affiliation{Division of Geological and Planetary Sciences, California Institute of Technology, Pasadena, CA 91125, USA}
\email{mattbel@caltech.edu}

\author[orcid=0000-0001-9665-8429]{Ian~Wong}
\altaffiliation{Authors contributed equally to this work.}
\affiliation{Space Telescope Science Institute, 3700 San Martin Drive, Baltimore, MD 21218, USA}
\email{iwong@stsci.edu}

\author[orcid=0000-0002-4950-6323]{Bryce~T.~Bolin}
\affiliation{Eureka Scientific, Oakland, CA 94602, USA}
\email{bbolin@eurekasci.com}

\author[orcid=0000-0002-7451-4704]{M.~Ryleigh~Davis}
\affiliation{Division of Geological and Planetary Sciences, California Institute of Technology, Pasadena, CA 91125, USA}
\email{rdavis@caltech.edu}

\author[orcid=0000-0003-2110-8152]{Steven~J.~Bromley}
\affiliation{Department of Physics, Auburn University, Leach Science Center, Auburn, AL 36849, USA}
\email{sjb0068@auburn.edu}

\author[orcid=0000-0002-9548-1526]{Carey~M.~Lisse}
\affiliation{Johns Hopkins University Applied Physics Laboratory, Planetary Exploration Group, Space Department, 11100 Johns Hopkins Road, Laurel, MD 20723, USA}
\email{carey.lisse@jhuapl.edu}

\author[orcid=0000-0002-8255-0545]{Michael~E.~Brown}
\affiliation{Division of Geological and Planetary Sciences, California Institute of Technology, Pasadena, CA 91125, USA}
\email{mbrown@caltech.edu}

\begin{abstract}
We present the first spectroscopic characterization of an interstellar object at mid-infrared wavelengths. Post-perihelion observations of 3I/ATLAS using the JWST/MIRI medium-resolution spectrometer were obtained on 2025 December 15--16 and 27 when the object was at heliocentric distances of 2.20 and 2.54~au, respectively. Our 5--28~\um\ spectra exhibit fluorescence features from several gaseous species, including the $\nu_2$ band of \water\ at 5.8--7.0~\um, the primary $\nu_2$ and associated hot bands of \cotwo\ around 15~\um, and a forbidden transition of atomic nickel at 7.507~\um. We also report the first direct detection of \methane\ in an interstellar object. The sharp rise in \methane\ production relative to \water\ suggests past \methane\ depletion from the outermost layers, with the observed \methane\ emerging from unprocessed subsurface material. Comparison of the volatile production rates measured during the two epochs indicate a significant reduction in overall outgassing over 12 days, with the measured \water\ activity level dropping more steeply than other species. As shown through near-nucleus coma mapping, 3I continues to display an extended source of \water\ production from icy grains entrained within the coma. Our production rate measurements confirm that 3I has a strongly enhanced \cotwo:\water\ mixing ratio relative to typical solar system comets, as well as a somewhat enriched \methane:\water\ value.
\end{abstract}


\section{Introduction}\label{sec:intro}
Interstellar objects (ISOs) are planetesimals that formed around other stars and were later ejected from their birth systems through dynamical interactions \citep[for a review, see][]{Fitzsimmons2024come,Jewitt2024ISOs}. During their brief transit through our Solar System, ISOs offer discrete glimpses into extrasolar small body populations and provide a valuable point of comparison for assessing commonalities and differences in planetesimal formation processes throughout the galaxy \citep[e.g.,][]{Gibson2025}.

The interstellar comet 3I/ATLAS (hereafter, 3I) is the third confirmed ISO after 1I/'Oumuamua and 2I/Borisov \citep{Denneau2025MPEC}, with an estimated nucleus cross-section diameter of 2.6~km \citep{Hui2026arXiv}. In contrast to 1I, which had an inactive appearance \citep{Meech2017Natur,Bolin2018}, 3I exhibits a significant extended coma \citep[e.g.,][]{Bolin2025MNRAS,Chandler2025arXiv,Santana-Ros2025,Seligman2025ApJL}. Its photometric evolution indicates that its coma consists of large, hundred micron sized dust grains \citep{Jewitt2025ApJL}. Ground-based observations \citep[][]{Belyakov2025RNAAS, delaFuenteMarcos2025A&A, Kareta2025ApJL, Opitom2025MNRAS, Yang2025ApJL} revealed that 3I is redder than 2I \citep{Jewitt2019ApJL,Bolin20202I} or typical cometary nuclei \citep{Lamy2004come}. Modeling of galactic dynamics suggests that 3I's excess velocity is consistent with a dynamical age ranging from 3 to 11~Gyr \citep{Hopkins2025ApJL,Taylor2025ApJL}.

A concerted effort has been undertaken to characterize the chemical makeup of 3I's coma. Ground-based spectroscopy at visible wavelengths yielded detections of gas-phase cyanogen and atomic nickel \citep{Hoogendam2025arXiv,Rahatgaonkar2025ApJL,SalazarManzano2025ApJL}, while radio observations by ALMA added methanol and hydrogen cyanide to the molecular inventory \citep{Roth2025arXiv, Coulson2026MNRAS}. Pre-perihelion space-based observations in the near-infrared with JWST and SPHEREx uncovered fluorescence signatures from H$_2$O, CO$_2$, and CO \citep{Cordiner2025,Lisse2025arXiv}. Post-perihelion SPHEREx measurements indicated a significant increase in CO production and the emergence of an additional emission feature at 3.2--3.4~\um\, likely due to organics \citep{Lisse2026}. Further evidence of 3I's evolving activity includes a bluing visible color \citep{Zhang2026} and the apparent asymmetry between its pre- vs. post-perihelion \water\ production trends \citep{Tan2026}. 

In this Letter, we present the first results from our JWST program targeting 3I at mid-infrared wavelengths during its outbound trajectory. Observations with the medium-resolution spectrometer (MRS; \citealt{wells2015,argyriou2023}) of the Mid-Infrared Instrument (MIRI) were obtained as part of Cycle 4 Director's Discretionary Time Program \#9442 (PI: M. Belyakov). In this work, we focus on characterizing the gas-phase component of 3I's coma through spectral modeling of the fluorescence bands and mapping of their spatial distribution.

\section{JWST/MIRI Observations} \label{sec:obs}

The JWST/MIRI MRS instrument consists of four integral field units (IFUs) that operate simultaneously and cover different portions of the MIRI wavelength range (referred to as channels). Three separate grating settings---short (A), medium (B), and long (C)---are needed to construct an uninterrupted spectrum from 5 to 28~\um. Two sets of A/B/C observations requiring six visits in total were included in the observing program to produce two complete mid-infrared spectra of 3I. All observations were executed on 2025 December 15--16. However, two of the observations (1 and 5) suffered from guide star acquisition failures and were repeated on 2025 December 27 with slightly reduced integration times as Observations 13 and 15. A summary of the observing circumstances and viewing geometries for the six successful visits to 3I is provided in \autoref{tab:obs} in the Appendix. For details on the data reduction and spectral extraction methodology, see \autoref{sec:methods}.

To improve spatial sampling of the coma and guard against cosmic-ray and other detector artifacts, each 3I observation consisted of four dithered exposures. Observations of a nearby empty background field were acquired immediately following each on-target observation. To avoid possible contamination from 3I's extended coma, the background field was placed $5'$ away from the target along a vector perpendicular to the sunward direction. The background observations employed a two-point dither pattern to optimize observing time usage, with the same per-dither integration time as the corresponding on-target observations. 

Due to the relatively uncertain orbital ephemeris of 3I at the time, the initial set of observations used a wide dither pattern to maximize on-sky coverage. Subsequent astrometry reduced the uncertainty ellipse, and the rescheduled observations on 2025 December 27 used a smaller dither pattern designed for extended objects.

\begin{figure*}[t!]
\centering
    \includegraphics[width=0.94\textwidth]{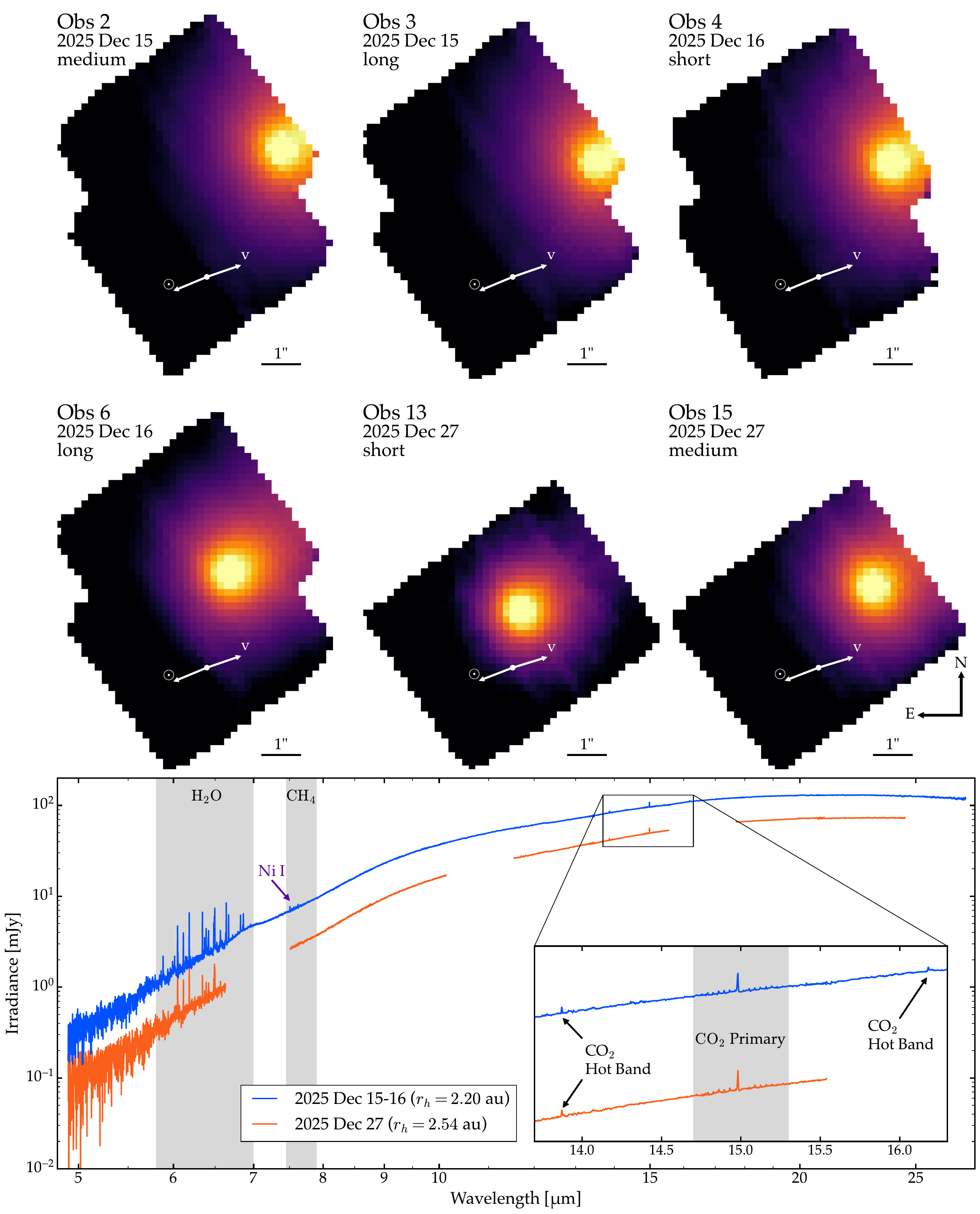}
    \caption{\textit{Top panels}: Median-stacked images derived from the fully calibrated Channel 2 data cubes for the six successful JWST/MIRI observations of 3I. A logarithmic stretch has been applied to accentuate the extended dust coma. The sunward and target velocity directions are indicated. The panels are labeled with the corresponding date and spectral grating setting. The first three observations suffered from poor pointing due to an uncertain ephemeris. There is low-level striping due to numerical artifacts from the IFU data cube building process.} \textit{Bottom panel:} Spectra of 3I extracted using a $1''$-diameter circular aperture. Observations 2, 4, and 6 (blue) are combined into a single spectrum by normalizing individual segments to align across their overlapping wavelengths. The spectra from Observations 13 and 15 (orange) are unaltered, illustrating the significant decrease in flux as the comet passed from a heliocentric distance of 2.20~au to 2.54~au. The main \water, \cotwo, \methane, and Ni fluorescence features are marked. The inset panel provides a zoomed-in view of the \cotwo\ primary and hot bands.
    \label{fig:fig1}
\end{figure*}

The uncalibrated data were processed using the \texttt{jwstspec} tool \citep{jwstspec} to produce fully calibrated, background-subtracted, and dither-combined data cubes. \autoref{fig:fig1} shows the wavelength-collapsed Channel 2 data cubes for the six successful observations of 3I. The target was situated close to the edge of the composite field of view in the first three observations. The off-center pointing proved to be problematic for Observation 3, where edge effects introduced significant correlated noise into the measured spectrum; as such, that observation was excluded from the analysis. The coma of 3I extends far beyond the spatial coverage of the MRS observations \citep{Lisse2026}. In these close-in views of the near-nucleus region, the dust continuum follows a mostly isotropic spatial distribution, with a slight bias towards the anti-sunward direction, similar to the anti-solar coma direction of 2I/Borisov \citep[][]{Bolin2020HST,Kim2020}. The extent of the coma's azimuthal asymmetry is quantified in \autoref{sec:azimuth}. Upon close inspection, some diagonal striping is evident in the images, which can be attributed to low-level numerical artifacts from the pixel resampling, dither combination, and spatial rectification processes required to construct the full IFU data cubes.

The bottom panel of \autoref{fig:fig1} shows the combined spectra of 3I from the two observing epochs: 2025 December 15--16 (blue; Observations 2, 4, 6) and 2025 December 27 (orange; Observations 13, 15). For these spectra, a fixed $1''$-diameter circular aperture centered on the target's centroid position was chosen to ensure that the extraction region was fully within the field of view across all observations. Adjacent spectral segments from different observations showed small flux offsets, which may stem from short-term variability in the brightness and activity level of the target. To produce a continuous spectrum, each individual segments was rescaled to align the flux levels within the overlapping wavelength regions. The systematic difference in flux level between the two epochs reflects the rapid dimming of 3I during its outbound trajectory. Based on the change in heliocentric and geocentric distance between the two observing epochs, the thermal flux of 3I is expected to have fallen by $\sim$40\%, which roughly matches the average decrease in the 8--25~\um\ irradiance between the spectra from the two epochs shown in \autoref{fig:fig1}.

\section{Fluorescence Modeling}\label{sec:volatiles}

The JWST/MIRI spectra of 3I reveal a rich collection of emission features from gas fluorescence within the coma. The Mid-infrared ro-vibrational bands of \water, \cotwo, and \methane, as well as the ground-state forbidden transition of monovalent atomic nickel (Ni I) can be seen above the thermal continuum (\autoref{fig:fig1}). For each volatile species, gas production rates $Q$ and rotational temperatures $\trot$ were derived from retrieval fits to continuum-subtracted spectra using the Planetary Spectrum Generator (PSG; \citealt{psg,villanueva2022}). Model spectra were generated according to the aperture size, viewing geometry, and target ephemeris at the midpoint time of the observations. A full description of the continuum removal and spectral retrieval methodology is provided in \autoref{sec:model}.

For each molecule, the first step in the analysis was a pixel-by-pixel retrieval of the column density (in units of m$^{-2}$) to check for regions with elevated optical depths ($\tau > 0.1$). Although PSG provides a first-order adjustment to the modeled fluorescence lines in such cases, this correction has only been benchmarked for small phase angles, while the JWST/MIRI observations were obtained at $\alpha > 18^{\circ}$. In order to ensure that the reported $Q$ and $\trot$ values are reliable, optically-thick regions near the nucleus were excluded from the analysis by using an annulus centered on the nucleus rather than a circular aperture.

The measured production rate depends on the assumed coma expansion velocity. For all models presented in this Letter, the expansion velocities were fixed to the values calculated from the scaling law of \citet{Ootsubo2012}, which gives $v \sim 540$ and 500~\ms\ for the two JWST/MIRI observing epochs, respectively. Pre-perihelion observations of 3I at similar heliocentric distances provided direct measurements of the expansion velocity, with values ranging widely from 240 to 510~\ms \citep{Roth2025arXiv,Coulson2026MNRAS}. Given that the production rate scales linearly with the expansion velocity, the roughly factor of two uncertainty in the expansion velocity at the time of the JWST/MIRI observations needs to be considered when interpreting the reported production rates. It is important to note, however, that the measured abundance ratios between the various gaseous molecular species are not affected by the assumed expansion velocity.

\subsection{Water}\label{subsec:h2o}

\begin{figure*}[t!]
\centering
    \includegraphics[width=\textwidth]{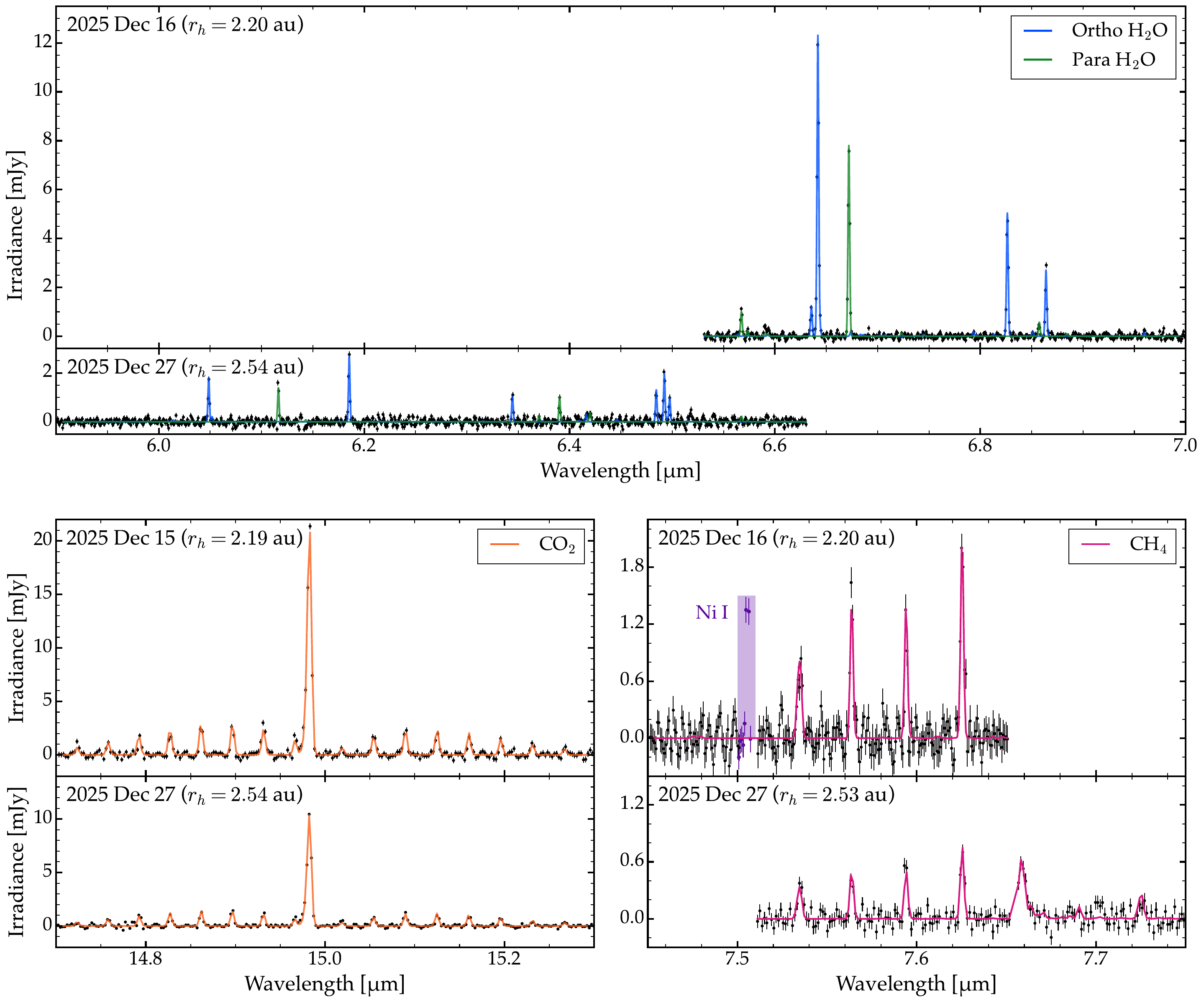}
    \caption{Top panels: continuum-subtracted JWST/MIRI spectra of 3I in the $\nu_2$ band of \water\ from both epochs, with the best-fit PSG coma fluorescence models overlaid. For the \water\ and \cotwo\ analyses, spectra were extracted using a 2--3$''$ annular aperture to avoid optically-thick regions, while for \methane, a circular $3''$-diameter aperture was applied. The plotted error bars are based on the measured uncertainties from the spectral extractions and have been scaled to yield a reduced $\chi^2$ of unity for the best-fit models.} The contributions from ortho- and para-\water\ are shown separately in blue and green, respectively. The line intensities are significantly lower in the second epoch, corresponding to a sizable decrease in \water\ production. Bottom left panels: analogous fits to the $\nu_2$ band of \cotwo. Bottom right panels: \methane\ fluorescence model and data. The emission feature highlighted in purple at 7.5066~\um\ is from the ground-state forbidden transition of Ni I. To illustrate the systematic decrease in production between the two epochs, the vertical scale is identical within each set of panels.
    \label{fig:fig2}
\end{figure*}

The mid-infrared contains prominent emission lines between 5.8 and 7.0~\um\ from the $\nu_2$ vibrational band of gas-phase \water. The $\nu_2$ manifold of \water\ spans the transition region between grating/channel settings 1B and 1C, which were sampled by Observations 2 and 6 on 2025 December 15--16; only the medium grating was used on 2025 December 27 (Observation 15). From the pixel-by-pixel analysis of Observations 2 and 6, optical depths of $\tau > 0.1$ were measured throughout the innermost $1\overset{''}{.}0$--$1\overset{''}{.}5$ circular region around the nucleus; in contrast, during the second epoch, the \water\ coma was optically thin throughout the field of view. Previous near-infrared observations found significant extended production of \water\ in the coma \citep[e.g.,][]{Cordiner2025,Lisse2026}. To obtain estimates of the effective \water\ production rate that more closely approximate the total production, while simultaneously avoiding optical depth effects in the immediate near-nucleus region, an annular spectral extraction aperture spanning a 2--3$''$ diameter was selected. This aperture was not applicable to Observation 2, given that the target was close to the edge of the field of view; as such, the corresponding spectrum was excluded from the \water\ fluorescence analysis.The continuum-removed spectra in the \water\ fluorescence region from Observations 6 and 15 are shown in the top panels of \autoref{fig:fig2}. More than 15 distinct emission peaks are clearly discernible above the noise level in the data. 

At the spectral resolution of the MRS ($R \sim 3000$), the ro-vibrational lines of the ortho and para nuclear-spin isomers are easily distinguishable, with the two isomers distinguished in blue and green respectively in \autoref{fig:fig2}. PSG provides separate molecular line lists for ortho- and para-\water, enabling direct measurement of the ortho-to-para ratio (OPR). As part of the pixel-by-pixel analysis, OPR values were obtained across the central region of the \water\ coma and were consistent with the equilibrium value of 3 when accounting for both measurement uncertainties and uncertainties stemming from the continuum removal process. When retrieving the overall \water\ production rates from the annulus-extracted spectra, the OPR was fixed to 3.

\begin{figure*}[t!]
\centering
    \includegraphics[width=\textwidth]{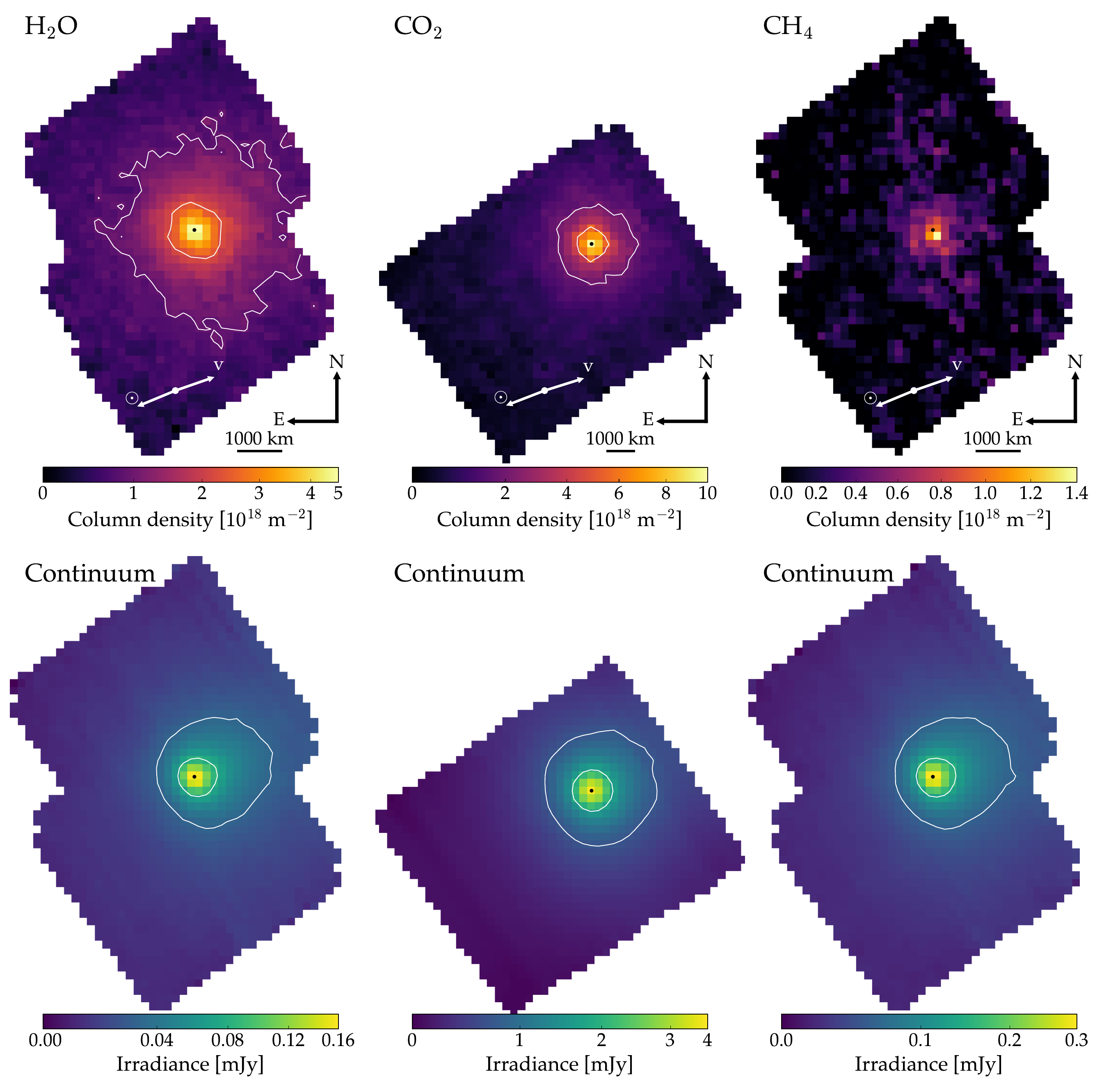}
    \caption{Top row: coma maps of \water, \cotwo, and \methane, computed as the molecular column density from a pixel-by-pixel retrieval analysis} for Observations 6, 15, and 6, respectively. The sunward and target velocity directions are denoted by the white arrows. The target centroids, computed as the photocenter in the median-stacked images (\autoref{fig:fig1}), are marked with the black points. For \water\ and \cotwo, the white contours correspond to emission levels of 50\% and 20\% relative to the maximum value and illustrate the slight anti-sunward extension of the respective comae. The contours are omitted in the \methane\ panel due to the poorer signal-to-noise ratio of the data. Bottom row: corresponding median-stacked images across the continuum wavelengths within each fluorescence region. The \water\ coma map shows significant enhancement in emission level at large nucleocentric distances relative to the dust continuum, indicating extended \water\ production within the coma.
    \label{fig:fig3}
\end{figure*}

For Observation 6, the measured \water\ production rate was $\qwater = (3.78 \pm 0.03) \times 10^{27}$~\persec, with a rotational temperature of $\trot = 16.9 \pm 0.3$~K. A significant reduction in \water\ production is evident in the second epoch (Observation 15). The PSG retrievals yielded $\qwater = (1.05 \pm 0.02) \times 10^{27}$~\persec and $\trot = 10 \pm 1$~K. The best-fit coma fluorescence models are shown in \autoref{fig:fig2}; the ortho- and para-\water\ contributions are plotted in different colors. The models provide an excellent match to the fluorescence peak amplitude distribution across the $\nu_2$ manifold in both observing epochs.

\autoref{fig:fig3} shows the spatial distribution of \water\ abundance in 3I's coma for Observation 6, which produced the strongest detection of the molecule. The \water\ coma is mostly isotropic, with a discernible extension in the anti-sunward direction relative to the target centroid. The full azimuthal profile of the \water\ coma is shown in \autoref{fig:fig7} in the Appendix. To facilitate examination of the relative extents of the dust and \water\ emission, the \water\ coma map is plotted alongside the continuum emission distribution within the \water\ fluorescence region. It is clear that the \water\ emission shows a significantly broader radial distribution than the dust, indicating that sublimation of \water\ ice from icy grains lofted into the coma by 3I's activity contributes significantly to the total gas-phase \water\ production \citep[e.g.,][]{Cordiner2025,Tan2026}.

\subsection{Carbon Dioxide}\label{subsec:co2}
JWST and SPHEREx observations of 3I have established \cotwo\ as the primary driver of its activity \citep{Cordiner2025,Lisse2025arXiv}, explaining the reported onset of outgassing at distances beyond $r_h = 6$~au \citep{Feinstein2025ApJL,Martinez-Palomera2025ApJL}. The new JWST/MIRI spectra present robust detections of the primary $\nu_2$ bending mode of \cotwo\ centered at 15~\um, along with the associated hot bands near 13.89 and 16.18~\um\ (\autoref{fig:fig2}). The continuum-subtracted spectra in the \cotwo\ primary band region, which were captured with the medium grating setting in Observations 2 and 15, are plotted in the bottom left panels of \autoref{fig:fig3}. The central Q-branch peak at 14.98~\um\ is flanked by several lines from the ro-vibrational P- and R-branches. These spectra were extracted from the same 2--3$''$ diameter annular aperture that was used for \water\ fluorescence modeling. The pixel-by-pixel examination revealed optical depths above 0.1 throughout the innermost $2''$-diameter region, rising as high as $\tau\sim0.5$ at the centroid pixel.

Retrievals in the primary \cotwo\ band at 14.7--15.3~\um\ yielded $\qcotwo = (8.70 \pm 0.09) \times 10^{27}$~\persec and $\trot = 46 \pm 1$~K for the 2025 December 15 observation, and $\qcotwo = (5.42 \pm 0.06) \times 10^{27}$~\persec and $\trot = 49 \pm 1$~K on 2025 December 27. The measured $\trot$ values are consistent across the two epochs, while being significantly lower than the reported $\trot$ of $\sim$90~K obtained from the previously published analysis of JWST/NIRSpec PRISM observations \citep{Cordiner2025}. At the low spectral resolution ($R\sim100$) of the JWST/NIRSpec PRISM spectrum, the individual transition lines within the P- and R-branches of the $\nu_3$ asymmetric stretch band at $\sim$4.3~\um\ are fully blended and cannot be isolated from the underlying continuum. The possible presence of an underlying \cotwo\ ice absorption band can greatly affect the continuum modeling across the fluorescence feature. Such an absorption band, if unaccounted for in the continuum subtraction, would artificially depress the band profile near the central Q-branch line, effectively broadening the distribution of emission strength across the P- and R-branches and yielding a positive bias in the $\trot$ estimate.Follow-up JWST/NIRSpec observations of 3I (Program \#5094, PI: M. Cordiner) using a higher-resolution grating setting (G395H; $R\sim2700$) were collected in 2025 December and will provide a more robust estimate of $\trot$ based on the $\nu_3$ band.


While the hot bands of \cotwo\ have not been previously reported in the literature on comets, they have been detected in recent observations of gas-rich protoplanetary disks with the JWST/MIRI MRS. Modeling of the $\nu_2$ \cotwo\ manifold observed in GW~Lup by \cite{Grant2023ApJL} demonstrated that detailed radiative transfer modeling within an optically thick regime is necessary to self-consistently reproduce the amplitudes of the primary and hot bands. Currently, there is no model available for \cotwo\ that combines cometary non-LTE radiative transfer processes as presented in the PSG with optically-thick radiative transfer modeling for the near-nucleus region. In the context of 3I, the presence of hot bands suggests that collisionally-induced pumping of high-energy ro-vibrational levels may be occurring in the dense innermost region of the \cotwo\ coma, given that fluorescence modeling with PSG is unable to simultaneously provide a satisfactory fit to the primary and hot \cotwo\ bands.

The \cotwo\ column density map measured from the second epoch (Observation 15) is plotted in \autoref{fig:fig3}. It is evident that \cotwo\ exhibits a much more compact coma than \water\ in the near-nucleus region. The radial profile of 3I at the \cotwo\ fluorescence wavelengths is slightly more compact than the dust across the full range of sampled nucleocentric distances, obviating the need to invoke a significant extended source within the coma. This dependence is in line with the findings of \citet{Cordiner2025}. From the 2025 August SPHEREx observations \citep{Lisse2025arXiv}, which probed regions far from the nucleus where optical depth is not a concern, the \cotwo\ coma is apparent at much farther distances ($\sim$3$'$) than the \water\ coma ($\sim$0.5$'$) due to the higher signal-to-noise ratio of the detected \cotwo\ feature, but follows a radial dependence that is steeper than the dust continuum's $1/\rho$ profile. As in the case of \water, the \cotwo\ coma shows a slight azimuthal asymmetry with higher abundance in the anti-sunward direction, though neither is as asymmetric as the dust, as shown in \autoref{fig:fig7}.

\subsection{Methane}\label{subsec:ch4}

Four fluorescence peaks of the $\nu_4$ triply degenerate bending mode of \methane\ are evident between 7.50 and 7.65~\um\ in both the sub-band 1C and 2A spectra obtained in Observations 6 and 13, respectively (see the bottom right panels of \autoref{fig:fig2}). This constitutes the first reported direct detection of \methane\ on any ISO. Near-infrared spectrophotometry of 3I collected by the SPHEREx spacecraft in 2025 December revealed a broad emission signature at 3.2--3.6~\um\ \citep{Lisse2026}---a range that contains the $\nu_1$ symmetric stretching band of \methane. However, SPHEREx lacks the requisite spectral resolving power to identify the specific hydrocarbon producing the spectral feature, as several other major cometary volatiles have fluorescence bands in that region (e.g., methanol, ethane, formaldehyde, and polycyclic aromatic hydrocarbons). 

\begin{figure}[t!]
\centering
    \includegraphics[width=\columnwidth]{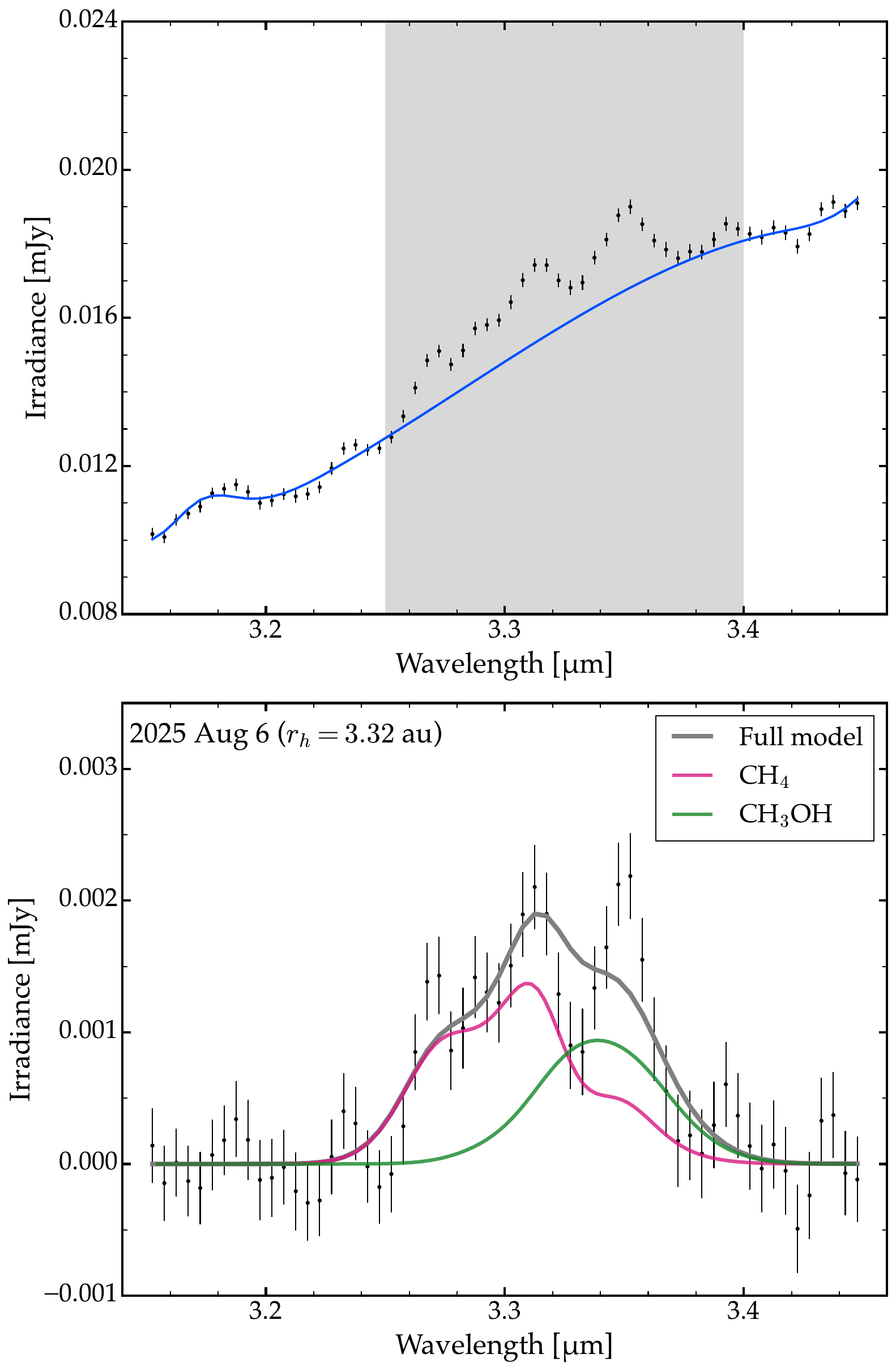}
    \caption{Top panel: the JWST/NIRSpec PRISM spectrum of 3I in the vicinity of the \methane\ $\nu_3$ fluorescence band (black points). The shaded region denotes the wavelength range that includes the detected emission bands, which was masked prior to continuum fitting. The blue curve is the cubic spline fit to the continuum points. Bottom panel: the continuum-subtracted spectrum (black points), with the best-fit PSG model overplotted. The gray curve corresponds to the full model, with the individual contributions from \methane\ and \methanol\ plotted in magenta and green, respectively.}
    \label{fig:fig4}
\end{figure}

The pixel-by-pixel modeling analysis revealed that 3I's \methane\ coma is optically thin across the field of view. A $3''$-diameter circular aperture was used to extract the spectra plotted in \autoref{fig:fig2} and estaimate the \methane\ production rate and rotational temperature. The PSG retrievals yielded $\qmethane = (4.2 \pm 0.2) \times 10^{26}$~\persec and $\trot = 37 \pm 2$~K for the first epoch. By the time of the second epoch, \methane\ production had fallen by almost 50\%, while the rotational temperature remained unchanged, following the same general trends seen for \cotwo: $\qmethane = (2.3 \pm 0.1) \times 10^{26}$~\persec and $\trot = 37 \pm 3$~K. The best-fit fluorescence model spectra are plotted in the bottom right panels of \autoref{fig:fig2}. Close inspection of the sub-band 2A spectrum reveals marginal detections of weaker emission features beyond 7.65~\um\ with amplitudes that are consistent with the model. Due to the low amplitude of the \methane\ fluorescence lines, which are at or below the level of noise in the individual pixel spectra, the corresponding coma abundance map from the first epoch (Observation 6) is noisy (\autoref{fig:fig3}). The spatial pixel with the highest \methane\ column density is situated to the southwest of the target centroid, which contrasts with the slight anti-sunward bias in the \water\ and \cotwo\ coma maps. However, while it may be tempting to attribute this difference in spatial extent to a distinct production region and/or mechanism, the exact morphology of the column density distribution in the near-nucleus region is poorly constrained and varies strongly when adjusting the details of the pixel-by-pixel continuum modeling. 

The detection of \methane\ in the JWST/MIRI observations raises the prospect of a previously unreported \methane\ fluorescence feature in the 2025 August 6 JWST/NIRSpec observations published by \citet{Cordiner2025}. In the near-infrared, the $\nu_3$ asymmetric stretching band of \methane\ is located near 3.3~\um. We reanalyzed the NIRSpec observations following a similar methodology to the MIRI data, and the irradiance spectrum was extracted using a $0.6''$-diameter circular aperture (see \autoref{sec:methods}). A notable feature in the dedicated background spectral images is the presence of a spatially uniform emission signature stemming from polycyclic aromatic hydrocarbons (PAHs) in galactic dust near 3.3~\um\ \citep[e.g.,][]{Tsumura2013PASJ}, which coincides with the region of \methane\ fluorescence. Because the amplitude of this feature varies across the sky, the background field was placed close to 3I ($3'$ away) to provide a reliable contemporaneous measurement of the PAH emission strength. Continuum removal in the 3.15--3.45~\um\ region revealed three positive bumps in the residual spectrum (top panel of \autoref{fig:fig4}). The first two features spanning 3.25--3.33~\um\ are well-matched by the combined profile of the P, Q, and R-branch fluorescence lines of \methane, which are fully blended at the low resolution of the PRISM spectrum. The final band centered at $\sim$3.35~\um\ is broadly consistent with both methanol (\methanol) and ethane (\ethane). Each of these molecules possesses additional weaker fluorescence features beyond 3.4~\um; however, the signal-to-noise ratio of the JWST/NIRSpec spectrum and the presence of significant correlated noise do not permit a reliable detection.


Coma modeling of the near-infrared fluorescence feature was carried out using PSG (see \autoref{sec:model}). Several compositions were considered, including \methane\ only, \methane+\methanol, \methane+\ethane, and \methane+\methanol+\ethane. The latter three mixed configurations provided similar fit quality and comparable $\qmethane$ values. Motivated by the previously reported detection of \methanol\ on 3I from radio observations \citep{Roth2025arXiv}, we present the results from the \methane+\methanol\ retrieval. The bottom panel of \autoref{fig:fig4} shows the best-fit PSG model alongside the continuum-subtracted JWST/NIRSpec spectrum. The retrieved production rates are $\qmethane = 1.4 \times 10^{25}$~\persec and $Q_{\rm CH_{3}OH} = 1.6 \times 10^{25}$~\persec. A systematic relative uncertainty of 30\% is placed on the production rate measurements based on the variance in values when altering the modeled composition and continuum subtraction methodology. We caution that the possible presence of an underlying absorption feature from solid hydrocarbons (e.g., \ethane, \methanol, and/or tholins) in the $\sim$3.3--3.4~\um\ region may cause an underestimation of the true \methane\ production rate from the JWST/NIRSpec PRISM spectrum---perhaps by up to a factor of a few. Future analysis of the JWST/NIRSpec observations obtained with the G395H grating will enable a more detailed exploration of the near-infrared fluorescence band inventory.

\subsection{Nickel}\label{subsec:Ni}
Previous detections of emission lines from electronic transitions from gaseous atomic nickel in distant cometary comae \citep{Manfroid2021Natur} and on 2I \citep{Guzik2021Natur} have raised the question of how transition metal species can appear as atomic gases at temperatures as low as 150~K---well below the sublimation temperature of Ni-bearing minerals or Ni metal. The short parent lifetime and $1/\rho$ spatial distribution of gaseous Ni have been invoked to suggest that atomic Ni is produced from photodissociation of organometallic complexes, e.g. Ni(CO)$_4$ \citep{Bromley2021PSJ,Guzik2021Natur,Hutsemekers2021A&A,Manfroid2021Natur}.

In the case of 3I, fluorescence from Ni I at near-UV and visible wavelengths has been reported by \citet{Hoogendam2025arXiv}, \citet{Hutsemekers2026}, and \citet{Rahatgaonkar2025ApJL}. Interestingly, while the Ni/Fe ratio was initially found to be extremely elevated from early pre-perihelion measurements \citep{Rahatgaonkar2025ApJL}, it has since steeply declined to Sun-like levels \citep{Hutsemekers2026}. However, the absolute production rates remain significantly higher than those observed on solar system comets.

The JWST/MIRI observations access ground-state transitions from the lowest-energy levels of atomic Ni. A query for forbidden electronic transitions of Ni I in the 5-28~\um\ range from the NIST Atomic Spectra Database \citep{NIST_ASD} returned six transitions. The 7.5066~\um\ 3d$^8$($^3$F)4s$^2$($^3$F$_3$) $\to$ 3d$^8$($^3$F)4s$^2$($^3$F$_4$) line is a transition from a 0.165 eV upper state to the ground state and is clearly detected in the 2025 December 16 spectrum, as shown in \autoref{fig:fig2}. The next brightest line, 3d$^8$($^3$F)4s$^2$($^3$F$_2$) $\to$ 3d$^8$($^3$F)4s$^2$($^3$F$_3$) at 11.307~\um, cascades into the upper level of the 7.5066~\um\ transition from a higher energy level of 0.275 eV. The fluorescence efficiency of the 11.307~\um\ line is roughly three times weaker than that of the 7.5066~\um\ line. Close inspection of the JWST/MIRI spectra uncovers the 11.307~\um\ line at approximately the expected flux ratio relative to the stronger 7.5066~\um\ line, given our computed fluorescence efficiencies. A [Ni I] line at 3.119~\um\ has a comparable fluorescence efficiency to the 7.5066~\um\ line and should be easily detected by JWST/NIRSpec observations with the medium- or high-resolution spectral grating, which are part of Program \#5094 (PI: M. Cordiner, \citealt{Cordiner_2026}).  Our observations do not provide a measurement of the Fe/Ni ratio, as there are no strong forbidden transitions from Fe I in the MRS wavelength range, except for the 24.04~\um\ line, which is in a region of the spectrum dominated by thermal emission and significant systematics that preclude its detection.

To compute Ni production rates from the two observed lines, fluorescence efficiencies (i.e., $g$-factors) were first calculated following the method described in \cite{Bromley2021PSJ}. Next, the number of Ni atoms within the spectral extraction aperture is given as
\begin{equation}
    N_{\text{atom}} = 4 \pi \Delta^2 \underbrace{\frac{ F_\nu d\lambda}{g \cdot h \cdot \lambda}}_\text{\# Density} \underbrace{\frac{1}{\pi \rho^2}}_\text{Aperture},
\end{equation}
where $F_\nu d\lambda$ is the integrated line flux in Jy$\cdot$\um, $\rho$ is the corresponding extraction radius in distance units, and $\Delta$ is the observer distance. Finally, the Haser model provides the absolute production rate \citep{Haser2020PSJ}:
\begin{equation}\label{eq:qni}
    \qnickel =  2 \pi \rho v N_{\text{atom}} \times  \frac{(\gamma_p-\gamma_d)/ \gamma_d}{\int\limits_0^{\rho/\gamma_p}K_0(x)dx-\int\limits_0^{\rho/\gamma_d}K_0(x)dx},
\end{equation}
where $\gamma_d = 9\times10^5$~km and $\gamma_p= 200$~km are the scale lengths of the daughter and parent in the hypothesized photodissociation reaction (i.e., Ni I and the putative nickel-bearing carbonyl molecule) taken from \cite{Bromley2021PSJ}, $K_0$ is the modified zeroth-order Bessel function of the second kind, and $v=0.8/\sqrt{r_h}=0.54$~km~\persec\ is the coma expansion velocity \citep{Ootsubo2012}. The resulting production rate is $\log{\qnickel} = 23.72 \pm 0.12$~\persec. Using a more straightforward $1/\rho$ fit without accounting for photodissociation of a parent species, which entails removing the final multiplicative factor containing the $\gamma$ terms from Equation~\eqref{eq:qni}, yields a slightly higher production rate of $\log{\qnickel} = 23.89 \pm 0.15$~\persec. The production rate estimated from the 11.307~\um\ line is slightly larger at $\log{\qnickel} = 24.0 \pm 0.2$~\persec and $24.2 \pm 0.2$~\persec\ for the two aforementioned models, respectively. The reported production rate uncertainties incorporate the spread in calculated band areas obtained from iterative Monte Carlo resampling of the data based on the measurement uncertainties line, as well as the uncertainty on the fluorescence efficiency (roughly 20\%, based on information from the NIST Atomic Spectra Database; \citealt{NIST_ASD}).

It is worth noting that all detections of atomic Ni in cometary nuclei to date have probed higher-energy transitions at near-UV wavelengths, in contrast with the low-energy transitions (0.11 and 0.165 eV) observed by JWST/MIRI. The population of the upper level of the two lines is likely to be primarily determined by fluorescence pumping of strong UV lines, which cascade down to the upper state of the 11.307~\um\ transition. Many of the Ni lines in the UV and visible region exhibit $\sim20\%$ uncertainties in their Einstein $A$ coefficients and corresponding absorption rates, suggesting that the $g$-factor for both the 7.5066~\um\ and 11.307~\um\ lines may be subject to additional uncertainties. Contemporaneous ground-based spectroscopy targeting the nickel lines in the near-UV would help confirm the validity of production rates measured in the mid-infrared. Despite the issues inherent to a production rate derived from two lines, our results are highly consistent with pre-perihelion measurements \citep{ Rahatgaonkar2025ApJL,Hutsemekers2026}.

\begin{figure*}[t!]
\centering
    \includegraphics[width=\textwidth]{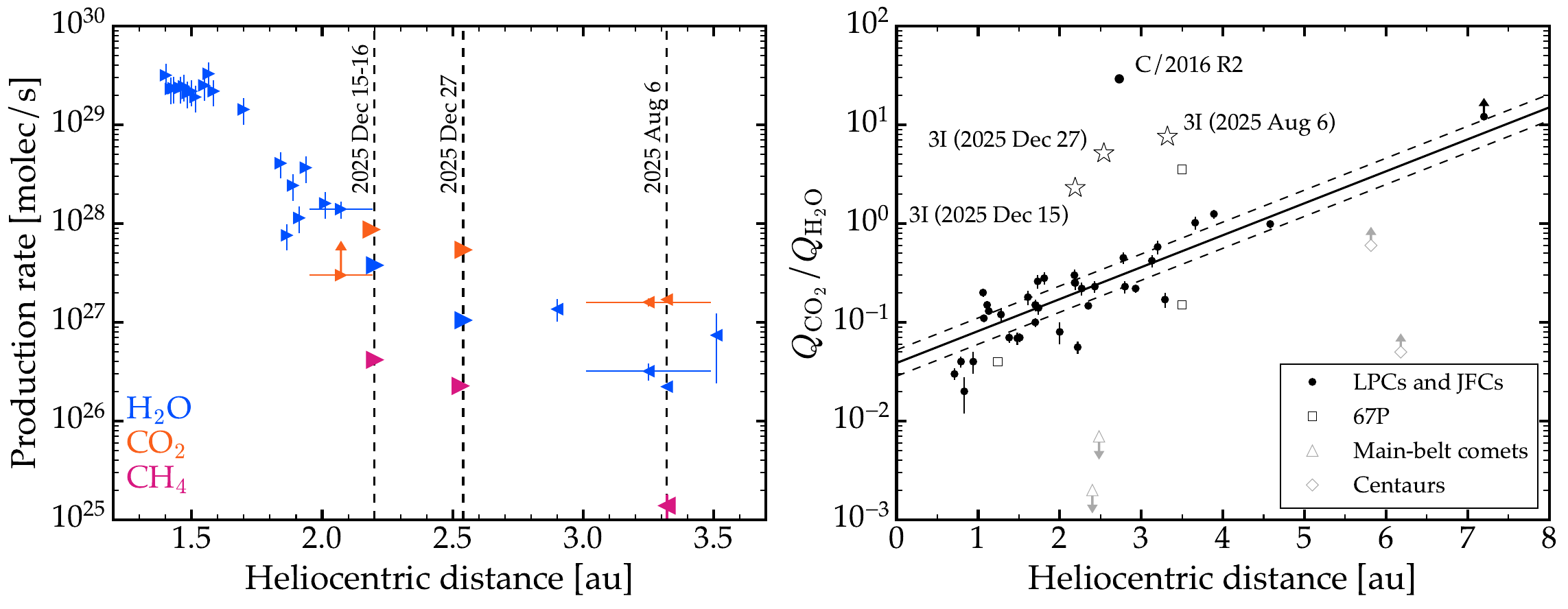}
    \caption{Left panel: a compilation of \water, \cotwo, and \methane\ production rates for 3I published in \citet{Combi2026}, \citet{Cordiner2025}, \citet{Lisse2025arXiv}, \citet{Xing2025ApJL}, \citet{Lisse2026}, and this work, plotted as a function of heliocentric distance. The JWST epochs (both NIRSpec and MIRI) are marked with vertical dashed lines. The MIRI measurements and the $\qmethane$ value obtained from the reanalysis of the NIRSpec observations are shown with larger symbols. Left-facing triangles denote pre-perihelion measurements, and vice versa. Right panel: \cotwo-to-\water\ production ratios versus heliocentric distance for various active objects, sorted by dynamical class, with 3I's measurements marked with the star symbols. This plot is a reproduction of Figure 6 in \citet{Cordiner2025} (see reference therein), with the new JWST/MIRI-derived values added. The solid and dashed black lines indicate the best-fit log-linear trend and associated $1\sigma$ bounds for the population of long-period comets (LPCs) and Jupiter-family comets (JFCs), excluding the exceptionally \cotwo-rich object C/2016~R2. The measured \cotwo:\water\ mixing ratios for 3I are significantly enriched relative to the trend, as is the case for C/2016~R2. See text for details.}
    \label{fig:fig5}
\end{figure*}

Within this context, we briefly examine the implications of the Ni detection on 3I. The common hypothesis is that Ni is trapped in organometallic complexes (either metal carbonyls, or metal-bonded polycyclic aromatic hydrocarbons; \citealt{Bromley2021PSJ,Manfroid2021Natur}), which produce bare Ni atoms upon photodissociation. Our production rate measurements are consistent with the pre-perihelion value obtained at roughly the same heliocentric distance: $\log{\qnickel} = 23.80 \pm 0.03$~\persec\ at 2.19~au \citep{Hutsemekers2026}. If these measurements are taken to imply symmetric inbound and outbound production rates, it would appear that the steep heliocentric distance dependence of Ni production does not stem from the comet's long passage through the interstellar medium. An alteration-driven production (e.g., via irradiation by galactic cosmic rays) would result in a distinct nickel production rate, as the outer surface of the comet was shed during its perihelion approach. The symmetric pre-~vs.~post-perihelion behavior of the production rate, however, confirms that the metal's production mechanism and source are distinct from canonical volatile release in comets, as suggested by \cite{Rahatgaonkar2025ApJL}. The JWST/MIRI production rates for \cotwo\ and \water\ suggest that 3I releases $2.5\times10^{28}$ oxygen atoms from the gas, which can be roughly doubled to account for oxygen in the silicates. This crude estimate for the atomic abundance of oxygen yields a Ni/O ratio of 0.002\%. Given that the solar abundance of nickel relative to oxygen is approximately 0.34\% \citep{Lodders2003ApJ}, the observed  Ni:\water\ and Ni:\cotwo\ ratios indicate that the measured gaseous Ni emission---as well as the hypothesized organometallic complexes from which it is sourced---represents only a fraction of a percent of 3I's total nickel reservoir.

\section{Discussion}\label{sec:disc}

From a scientific standpoint, the rescheduling of Observations 13 and 15 proved fortuitous, as the 12-day baseline between the two epochs enables a self-consistent assessment of 3I's evolving gas production. Between 2025 December 15--16 and 27, 3I passed from 2.20 to 2.54~au (\autoref{tab:obs}), providing two distinct views of the object's volatile production at heliocentric distances not sampled by previous observations. The \water, \cotwo, and \methane\ production rates measured from the JWST/MIRI spectra are plotted in the left panel of \autoref{fig:fig5}, where the marked decrease in activity between the two epochs is clearly seen. Notably, the $\qwater$ values exhibit a significantly steeper drop between $r_h = 2.20$ and 2.54~au than the other volatiles, which can be attributed to the lower volatility of \water\ compared to \cotwo\ and \methane. The \water\ ice line in the Solar System is located around 2.5~au, and as 3I approached those heliocentric distances in 2025 December, \water\ production from the coldest regions of 3I's surface and coma was starting to shut down. Meanwhile, because of their much lower vapor pressures, \cotwo\ and \methane\ are expected to have remained fully activated.

\autoref{fig:fig5} shows a collection of published volatile production rates for 3I. The pre-perihelion data points include $\qwater$ and $\qcotwo$ values calculated from JWST/NIRSpec PRISM observations \citep{Cordiner2025}, $\qwater$ and $\qcotwo$ production rates obtained from SPHEREx spectrophotometry spanning $r_h=3.01$--3.49~au \citep{Lisse2025arXiv}, $\qwater$ measurements derived from observations by the Swift Observatory \citep{Xing2025ApJL}, and the $\qmethane$ value estimated from our reanalysis of the JWST/NIRSpec observations (Section~\ref{subsec:ch4}). Post-perihelion production rates include 18 $\qwater$ estimates derived from Lyman-$\alpha$ measurements by the Solar and Heliosphere Observatory (SOHO; \citealt{Combi2026}) and $\qwater$ and $\qcotwo$ values obtained from 2025 December SPHEREx observations spanning $r_h=1.95$--2.19~au \citep{Lisse2026}. Note that due to the optical thickness of the innermost \cotwo\ coma during the latest SPHEREx visit, which was not accounted for in the \cotwo\ activity measurement, the corresponding $\qcotwo$ value is a strict lower limit to the true production rate.

The JWST/MIRI-derived $\qwater$ values presented in this Letter continue the steep drop in \water\ production indicated by the other post-perihelion measurements. Notably, both the SOHO and SPHEREx $\qwater$ values were obtained using large apertures ($> 10^5$~km) that are well-suited to capture the full extent of distributed \water\ production within 3I's expansive coma. It follows that, given the smooth heliocentric distance trend traced by all post-perihelion \water\ production rates, there is no indication that the $\qwater$ values retrieved from the MRS spectra are significantly underestimating 3I's total \water\ production.

The previous JWST/NIRSpec observations obtained at $r_h = 3.32$~au found that 3I is unusually rich in \cotwo\ relative to \water, with a measured $\qcotwo/\qwater$ ratio of $7.3\pm0.6$ \citep{Cordiner2025}. SPHEREx measurements from 2025 August likewise indicated an enhanced \cotwo:\water\ ratio of $5.0\pm1.3$ \citep{Lisse2025arXiv}. We report $\qcotwo/\qwater= 2.30\pm0.03$ and $5.16\pm0.13$ for the two epochs of JWST/MIRI spectroscopy, respectively. The right panel of \autoref{fig:fig5} shows an ensemble of published $\qcotwo/\qwater$ measurements for active objects (after \citealt{Cordiner2025}, adding our new mixing ratio measurements). While the heliocentric dependence of $\qcotwo/\qwater$ for 3I is roughly similar to that of typical long-period and Jupiter-family comets, 3I's mixing ratio measurements are enhanced by over an order of magnitude relative to most other published values obtained at similar heliocentric distances and approach the \cotwo:\water\ mixing ratio of the exceptionally volatile-rich comet C/2016~R2. 

The most notable finding from our mid-infrared spectroscopy of 3I is the robust detection of \methane\ production. As with \cotwo:\water, the \methane:\water\ mixing ratio of 3I may also be enhanced relative to typical solar system comets. However, previous measurements of the \methane\ production rate for solar system comets beyond $r_h = 2$~au are sparse. Excluding the exceptionally hypervolatile-rich comet C/2016~R2, which has a reported \methane:\water\ mixing ratio of $181\%\pm25\%$ \citep{McKay2019}, the handful of published $\qmethane/\qwater$ values spans 0.1--10\% \citep{LeRoy2015,DelloRusso2016,Bonev2017}. The $\qmethane/\qwater$ values for 3I from the two JWST/MIRI epochs are $11.0\% \pm 0.5\%$ and $21.6\% \pm 1.3\%$, respectively, with the higher mixing ratio in the second epoch consistent with the aforementioned pronounced reduction in \water\ outgassing as 3I approached the \water\ sublimation line during its outbound track. Contemporaneous SPHEREx observations on 2025 December 7--15 yielded a \methane:\water\ mixing ratio of $\sim$14\%, assuming that the emission from organics at 3.2--2.5~\um\ is dominated by \methane\ \citep{Lisse2026}. The SPHEREx measurement is interpreted as an upper limit for the mixing ratio, as the presence of \methanol, which was demonstrated from ALMA measurements \citep{Roth2025arXiv}, also contributes to the 3.4~\um\ excess. The agreement between the SPHEREx value, obtained from a large aperture, and the JWST/MIRI measurements suggests that any potential underestimation of $\qwater$ due to the limited MRS field of view does not have a substantial impact on the reported \methane:\water\ mixing ratio.

A peculiar feature of 3I's \methane\ activity history is the pre-perihelion $\qmethane$ value measured from the 2025 August 6 JWST/NIRSpec observation, which is almost an order of magnitude lower than the extrapolated trend line from the two post-perihelion measurements (\autoref{fig:fig5}). Meanwhile, the production rate of \cotwo\ displays monotonic combined pre- and post-perihelion activity modulation. The distinctive behavior of \methane\ indicates that a significant enhancement in \methane\ production occurred sometime after the 2025 August 6 pre-perihelion JWST/NIRSpec observation and prior to the post-perihelion JWST/MIRI visits.

Solid-phase \methane\ is hypervolatile, with a significantly lower sublimation temperature than \cotwo. Thermal modeling of airless bodies \citep[e.g.,][]{schaller2007,lisse2021,lisse2022} has demonstrated that \methane\ ice at or near the surface of 3I would have been vigorously sublimating at the time of the first reports of cometary activity during 3I's pre-perihelion track ($r_h\sim 6$~au; \citealt{Feinstein2025ApJL,Martinez-Palomera2025ApJL,Ye2025ApJL}). However, the \methane\ production rate measurements indicate that the molecule was not fully activated when 3I was inbound at $\sim$3.3~au. This could imply that 3I previously underwent a period of significant heating within its natal planetary system prior to its ejection into the cold interstellar medium \citep{Lisse2026} that depleted the \methane\ in the outermost layers. Consequently, the surviving reservoir of primordial \methane\ ice resides at depth and was only fully activated after the thermal wave induced by 3I's perihelion passage propagated into the interior. This behavior is mirrored by CO---even more volatile than \methane---which displayed a roughly 40-fold increase in production relative to \cotwo\ in the 2025 December SPHEREx observations of 3I when compared to pre-perihelion measurements \citep{Lisse2026}. An additional contributing factor to the increased production of these hypervolatile species could be the removal of the outermost layers of the surface through sublimation-driven mass loss during 3I's perihelion passage.

\begin{deluxetable*}{cccccccccccccccc}[t!]
\setlength{\tabcolsep}{0.65pt}
\tablewidth{0pc}
\renewcommand{\arraystretch}{0.9}
\tabletypesize{\small}
\tablecaption{
    JWST/MIRI Observation Details and Viewing Geometries
\label{tab:obs}}
\tablehead{
    Obs. Numb. & $\qquad\quad$ & UT Midpoint Time & $\qquad\quad$ & Grating Setting & $\qquad\quad$ & Exp. Time & $\qquad\quad$ & $V$ & $\qquad\quad$ & $\Delta$  & $\qquad\quad$ & $r_h$ & $\qquad\quad$ & $\alpha$  \\
    & & & & & & (s) & & (mag) & & (au) & & (au) & & (deg) 
}
\startdata
2 & & 2025 Dec 15 17:22:40 & & medium (B) & & 833 & & 15.31 & & 1.800 & & 2.189 & & 26.68 \\
3 & & 2025 Dec 15 12:49:14 & & long (C) & & 833 & & 15.30 & & 1.801 & & 2.181 & & 26.83 \\
4 & & 2025 Dec 16 00:49:25 & & short (A) & & 833 & & 15.31 & & 1.800 & & 2.195 & & 26.56 \\
6 & & 2025 Dec 16 06:22:11 & & long (C) & & 833 & & 15.32 & & 1.799 & & 2.202 & & 26.43 \\
\hline
13 & & 2025 Dec 27 11:20:29 & & short (A) & & 777 & & 15.62 & & 1.824 & & 2.530 & & 18.75 \\
15 & & 2025 Dec 27 20:20:30 & & medium (B) & & 777 & & 15.63 & & 1.826 & & 2.541 & & 18.46 \\
\enddata
\footnotesize{
\vspace{6pt}
\noindent \textbf{Note.} The reported exposure time is the sum across the four dithered exposures. The variables $V$, $\Delta$, $r_h$, and $\alpha$ are the apparent $V$-band brightness, distance from JWST, heliocentric distance, and phase angle of 3I at the midpoint of the observation as computed by JPL Horizons.}
\vspace{-0.5cm}
\end{deluxetable*}

\begin{acknowledgments}
This work is based on observations made with the NASA/ESA/CSA James Webb Space Telescope. The data were obtained from the Mikulski Archive for Space Telescopes at the Space Telescope Science Institute, which is operated by the Association of Universities for Research in Astronomy, Inc., under NASA contract NAS 5-03127 for JWST. The JWST/MIRI observations are associated with Program \#9442. The specific observations analyzed can be accessed via \dataset[doi:10.17909/p4qv-4p68]{https://doi.org/10.17909/p4qv-4p68}. We thank Geoff Blake, Carl Schmidt, Zachariah Milby, and Sierra L. Grant for enlightening discussions on spectroscopy. We extend a special thanks to Davide Farnocchia and Marco Micheli for updating the ephemerides of 3I/ATLAS, which allowed these observations to succeed. We also acknowledge Geronimo Villanueva, whose assistance on PSG greatly facilitated the pixel-by-pixel modeling work described in this Letter.
\end{acknowledgments}

\begin{contribution}
M. Belyakov conceived the project, wrote the JWST proposal, ensured acquisition of the observations, and led the modeling of nickel production and azimuthal profile fitting. I. Wong led the data reduction and primary PSG retrievals. M. Belyakov and I. Wong were equally responsible for writing the manuscript and coma mapping. B.T. Bolin helped design the observations, led supporting observations that enabled the JWST observations, and contributed to the text. M.R. Davis contributed to initial data reduction and writing of the manuscript. S.J. Bromley provided fluorescence efficiencies and contributed to the interpretation of nickel production. C.M. Lisse and M.E. Brown contributed to the data analysis and interpretation.
\end{contribution}
\appendix

\section{Data Reduction and Spectral Extraction}\label{sec:methods}

\autoref{tab:obs} summarizes the exposure details and viewing geometries for the six successful JWST/MIRI MRS observations of 3I collected as part of Program \#9442. The apparent magnitude and target distance information are provided for the midpoint time of each set of dithered exposures.

Data reduction and spectral extraction were performed following standard procedures using the publicly-available JWST data analysis tool \texttt{jwstspec} \citep[v0.9;][]{jwstspec}. This data processing framework has previously been applied to numerous JWST observations of small bodies, and detailed descriptions of the pipeline's functionality can be found in several published works \citep[e.g.,][]{rivkin2023,emery2024,wong2024}. First, the uncalibrated stacks of nondestructive detector reads were downloaded from the Mikulski Archive for Space Telescopes and passed through all three stages of the official JWST calibration pipeline to produce dark-corrected, flat-fielded, flux-calibrated, spatially-rectified, background-subtracted, and dither-combined data cubes. Fully calibrated data cubes were also generated for each dithered exposure to aid in outlier removal and vet the various molecular signatures detected in the spectra. The pipeline products analyzed in this work were generated using Version 1.20.2 of the JWST calibration pipeline \citep{jwst}, with requisite reference files drawn from context \texttt{jwst\_1471.pmap} of the JWST Calibration Reference Data System. 

Several custom settings that are not part of the default processing workflow were applied to optimize performance: (1) the cosmic ray detection threshold in the \texttt{jump} step in Stage 1 of the calibration pipeline was raised to $10\sigma$ to prevent spurious flagging of points in light of the intrinsically variable nature of 3I and possible imperfect non-sidereal tracking of the object during each exposure due to ephemeris uncertainties; (2) the \texttt{clean\_showers} routine was activated as part of the \texttt{straylight} step in Stage 2, which has been shown to greatly mitigate the dispersed flux (so-called showers) that arises from large cosmic ray impacts \citep{regan2024}; (3) the \texttt{residual\_fringe} step was performed in Stage 2 to improve the defringing of the flux-calibrated detector images prior to cube building; (4) the \texttt{outlier\_detection} step was skipped in Stage 3 to prevent erroneous pixel masking during the dither combination process that can often occur with variable moving object targets.

Next, the target centroid was determined by median-averaging the data cube along the wavelength axis and fitting a 2D Gaussian to the flux values within a $5\times5$ cutout region centered on the brightest pixel in the field of view. Spectra were extracted using various circular and annular apertures centered on the centroid position, with the sizes and extents chosen to avoid optically-thick regions. The angular sizes were converted to pixel units using the corresponding pixel scale of each channel: $0\overset{''}{.}13$ in Channel 1, $0\overset{''}{.}17$ in Channel 2, $0\overset{''}{.}20$ in Channel 3, and $0\overset{''}{.}35$ in Channel 4.

The \texttt{jwstspec} tool enables the user to customize the level of outlier rejection in the data cubes prior to spectral extraction. Each pixel flux array within the spectral extraction region is smoothed using a cubic spline fit, and points that exceed a certain multiple of the flux uncertainty $\sigma$ relative to the spline model are masked as outliers. Given the presence of narrow, large-amplitude emission features in the spectrum of 3I, the outlier rejection threshold was set to a very high value ($100\sigma$) to prevent unintentional removal of fluorescence peaks. Likewise, the post-extraction spectral outlier removal threshold, which is applied to a 21-pixel-wide moving median filter, was also set to $100\sigma$ to ensure that only extreme non-astrophysical outlier points were excluded. 

Remaining outliers in the extracted spectra were manually trimmed in \texttt{jwstspec} through careful comparisons with coma fluorescence models (see \autoref{sec:model}). PSG model spectra were generated for all common cometary volatile species with major emission bands in the mid-infrared, including \water, \cotwo, and various aliphatic organic species (e.g., \methane, \ethane). After identifying the locations of the molecular fluorescence lines, visible outliers were excised across the wavelength regions where no astrophysical spectral features are expected to occur. Lastly, as a final defringing step, the \texttt{ifu\_rfcorr} routine contained within the \texttt{extract\_1d} step of the JWST calibration pipeline was run on the outlier-removed spectra to fit and remove remaining periodic flux variations.

For the reanalysis of the UT 2025 August 6 JWST/NIRSpec IFU observations of 3I from Program \#5094 (PI: M. Cordiner), the data processing workflow was largely identical to the one described above. The on-target exposures (Observation 1) were paired with a set of dedicated background exposures (Observation 2), and pixel-by-pixel background subtraction was carried out as part of Stage 2 of the calibration pipeline. Unlike MIRI, the NIRSpec detectors do not exhibit discernible fringing. Instead, readnoise artifacts, which manifest as vertical striping in the raw detector images, are present and were removed by activating the \texttt{nsclean} step during the pipeline processing. A 3-pixel-radius ($0\overset{''}{.}6$ diameter) circular aperture was used to extract the spectrum.

\section{Spectral Modeling}\label{sec:model}
The PSG \citep{psg} was used to retrieve the abundances (parametrized as either nuclear production rate or local column density) and rotational temperatures of the volatile molecular species in 3I's coma. The retrieval tool includes a Monte Carlo sampler to calculate the posterior distributions and corresponding $1\sigma$ uncertainties for all parameters, which are scaled to ensure a reduced $\chi^2$ value of unity for the best-fit model. To generate the synthetic spectra, the PSG implements the Cometary Emission Model \citep{villanueva2022}, which accounts for the expansion and evolution of gas molecules through the coma, relevant photodissociation lifetimes, and fluorescence pumping rates within a fully non-local thermal equilibrium (non-LTE) framework. The line lists for all modeled species were drawn from the GSFC Fluorescence Database \citep[e.g.,][]{villanueva2011}. The coma expansion velocity at each epoch was set according to the scaling law of \citet{Ootsubo2012}: $v = 0.8 / \sqrt{r_h}$~km~\persec. The fluorescence modeling tool flagged instances where the optical depth within the modeled region exceeds $\tau=0.1$, and those pixels were excluded from further analysis.

The viewing geometry (heliocentric distance, observer distance, phase angle) at the midpoint of each observation was queried from JPL Horizons. By default, the wavelength solution contained within the JWST/MIRI data cubes is adjusted to match the target's stationary frame, and so the radial velocity relative to the observer was set to zero. Imperfections in the wavelength calibration were addressed by allowing for small shifts to the wavelength grid in order to align the model spectra with the observed emission lines. The required shifts were less than 1--2~nm in all cases---comparable to or smaller than the spectral resolution element of the MRS gratings. The synthetic spectra were downsampled to match the local spectral resolution of the modeled data using a Gaussian convolution kernel with a full-width half-maximum (FWHM) scale set by the empirical laws of \citet{pontoppidan2024} and \citet{banzatti2025}. In the case of the primary \cotwo\ band at 15~\um, the calculated FWHM of 0.0052~\um\ was found to be too large, resulting in a significant mismatch between the observed and modeled line shape of the Q-branch transition. A FWHM of 0.0045~\um\ provided a much better fit to the line profile.

As part of the fluorescence analysis of the JWST/MIRI observations, PSG retrievals were carried out on both aperture-extracted spectra and individual pixel spectra. To prepare the data for the retrievals, the spectra were trimmed to the fluorescence band regions and continuum-subtracted following a two-step process. First, a crude continuum removal was carried out by fitting a cubic polynomial to the aperture-extracted spectra. A preliminary set of retrieval runs was carried out on these spectra, and the corresponding best-fit coma fluorescence models were used to identify the locations of emission lines: all wavelengths with modeled emission amplitudes that exceed a certain fraction of the maximum value within a given band were masked. The masking threshold was set individually for different molecules and observations, pursuant to the characteristic line amplitude distribution and signal-to-noise ratio of the data. For \water, a threshold of 0.2\% was applied to Observations 2 and 6, while a higher cutoff of 0.5\% was used for the noisier Observation 15 spectrum. For \cotwo\ and \methane, a limit of 1\% was imposed across all observations.

\begin{figure}[t!]
\centering
    \includegraphics[width=\columnwidth]{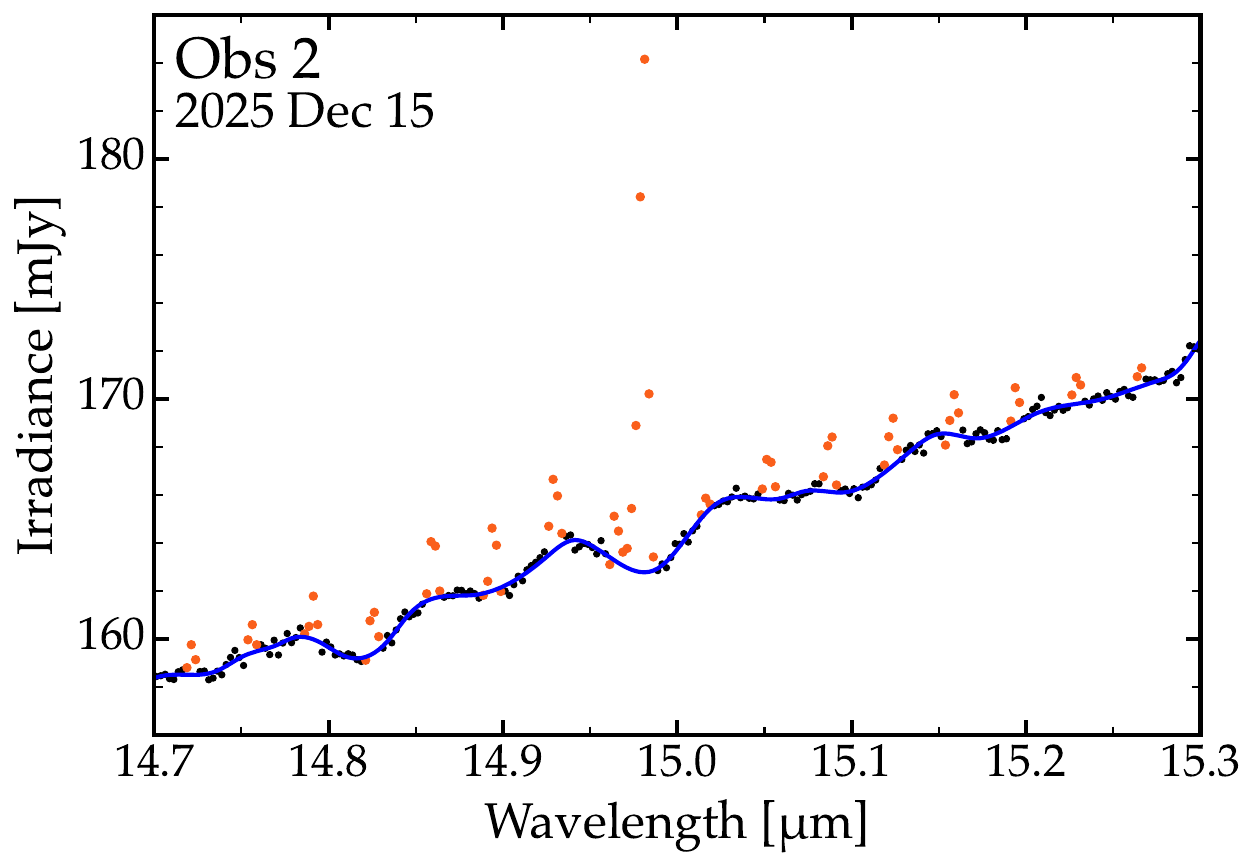}
    \caption{An example of the continuum modeling procedure applied to prepare the extracted JWST/MIRI spectra for retrieval analysis. The data points are the irradiance spectrum in sub-band 3B obtained from Observation 2 using a 2--3$''$ annular aperture. The orange points, which denote the wavelengths that contain \cotwo\ fluorescence emission, here masked prior to continuum modeling. The blue curve is the cubic spline fit to the continuum wavelengths, with the smoothing threshold set to $2.5n$, where $n$ is the number of continuum data points. See text for more details.}
    \label{fig:fig6}
\end{figure}

The second, more refined continuum removal step involved an error-weighted cubic spline fit to the masked spectra. The \texttt{scipy.interpolate.UnivariateSpline} function includes a smoothing threshold $s = \chi^2$ that controls the performance of the spline fitting. For each modeled fluorescence band, $s$ was adjusted via visual inspection to ensure that the continuum model adequately captures the large-scale modulations in flux without overfitting the point-to-point jitter. The optimal $s$ values ranged from $2n$ to $5n$, where $n$ is the number of continuum data points within the modeled spectral segment. The spline model was subtracted from the unmasked data to yield the continuum-corrected spectra shown in \autoref{fig:fig2}. \autoref{fig:fig6} provides an illustrative example of the continuum modeling for the \cotwo\ primary band detected in the first epoch (Observation 2).

Due to the lower spectral resolution of the JWST/NIRSpec PRISM observations, with individual fluorescence lines fully blended, the continuum removal procedure was much simpler. The extracted spectrum was trimmed to 3.15--3.45~\um\ and the region containing the aliphatic emission features (3.25--3.40~\um) was masked prior continuum modeling with a cubic spline; the smoothing threshold was set to $1.5n$. No wavelength shift was allowed when running the PSG retrievals on the NIRSpec spectrum.

\section{Azimuthal Profiles}\label{sec:azimuth}
\begin{figure}[t!]
    \includegraphics[width=\columnwidth]{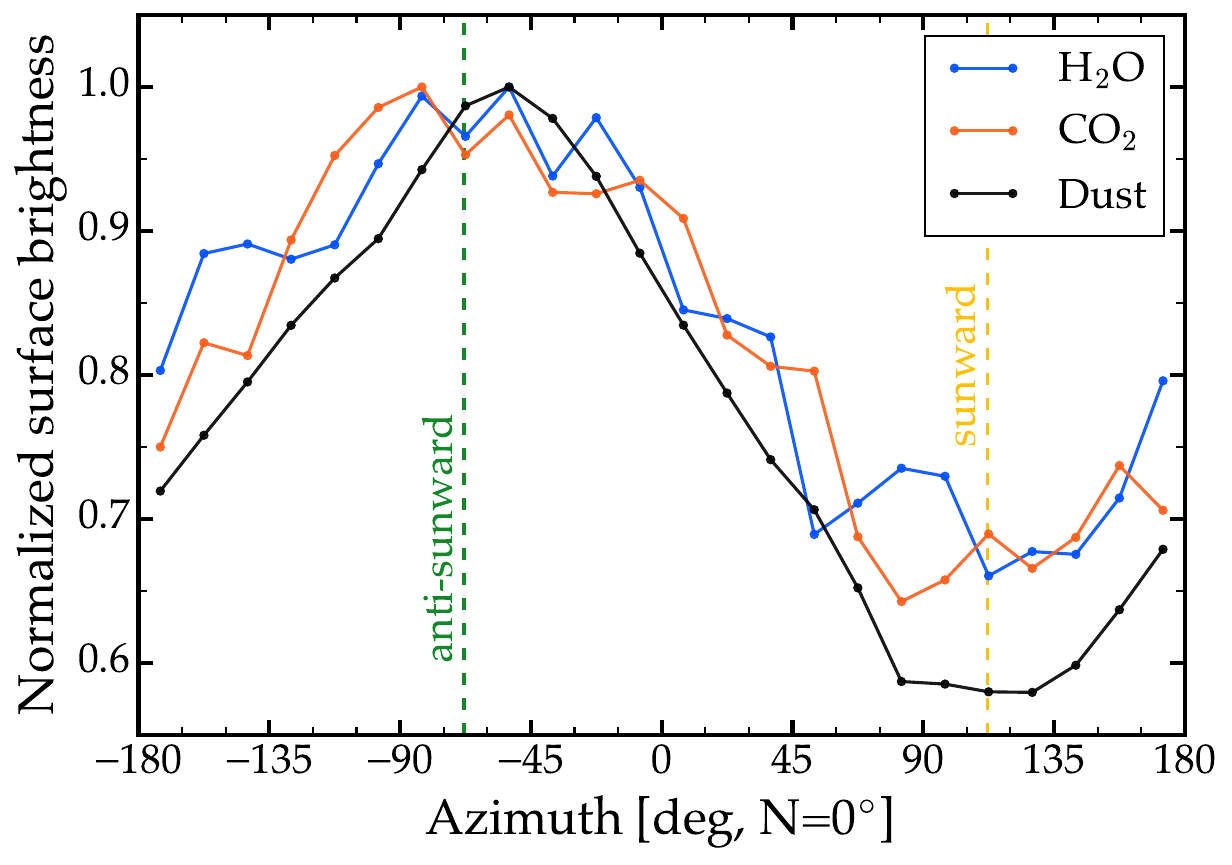}
    \caption{Azimuthal profiles of \water\ (blue) and \cotwo\ (orange) at $0\overset{''}{.}8$--$1\overset{''}{.}5$ from the optocenter, derived from Observations 6 and 15. The profile of the local dust continuum within the \cotwo\ primary band is also shown. The azimuth increases eastward with north at $0^\circ$. The sunward and anti-sunward directions are indicated by the yellow and green vertical dashed lines, respectively.}
    \label{fig:fig7}
\end{figure}

To quantify the spatial behavior of 3I's coma, azimuthal profiles of \water, \cotwo were obtained from Observations 6 and 15, respectively. For comparison, a dust profile was derived from the local continuum wavelengths within the \cotwo\ primary band in Observation 15. The profiles were computed from an annular aperture centered on the target centroid (see \autoref{sec:methods}) spanning radii from $0\overset{''}{.}8$ to $1\overset{''}{.}5$. This aperture was chosen to exclude both optically-thick pixels near the nucleus and regions outside of the combined field of view. The annulus was divided into 15$^\circ$-wide sectors to produce the azimuthal profiles, which were then normalized to unity at their maximum values. The measured \water, \cotwo, and dust continuum azimuthal profiles are shown in \autoref{fig:fig7}. All three profiles are visibly asymmetric in the near-nucleus region, with the strongest emission roughly aligned with the anti-sunward direction. The \water\ and \cotwo\ profiles exhibit a slightly more muted asymmetry than the dust continuum.

\facilities{JWST/MIRI.}
\software{\texttt{astropy} \citep{astropy2013,astropy2018,astropy2022}, \texttt{jwst} \citep{jwst}, \texttt{jwstspec} \citep{jwstspec}, \texttt{matplotlib} \citep{matplotlib}, \texttt{numpy} \citep{numpy}, PSG \citep{psg}, \texttt{scipy} \citep{scipy}.}

\bibliography{main}{}
\bibliographystyle{aasjournal}

\end{document}